\documentclass[pdflatex,sn-mathphys-num]{sn-jnl}

\usepackage{graphicx}%
\usepackage{multirow}%
\usepackage{amsmath,amssymb,amsfonts}%
\usepackage{amsthm}%
\usepackage{mathrsfs}%
\usepackage[title]{appendix}%
\usepackage{xcolor}%
\usepackage{float}
\usepackage{textcomp}%
\usepackage{manyfoot}%
\usepackage{booktabs}%
\usepackage{algorithm}%
\usepackage{algorithmicx}%
\usepackage{algpseudocode}%
\usepackage{listings}%

\theoremstyle{thmstyleone}%
\theoremstyle{thmstyletwo}%

\theoremstyle{thmstylethree}%

\begin{document}

\title[Correlated electronic structure and local spin in lead-copper-vanadium-bromine apatite: a DMFT study]{Correlated electronic structure and local spin in lead-copper-vanadium-bromine apatite: a DMFT study}


\author*[1]{\fnm{Ihor} \sur{Sukhenko}}\email{isukhenko@imp.kiev.ua}

\author[1]{\fnm{Volodymyr} \sur{Karbivskyy}}\email{karb000@ukr.net}


\affil*[1]{\orgdiv{Department of Physics of Nanostuctures}, \orgname{G.~V. Kurdyumov Institute for Metal Physics, NAS of Ukraine}, \orgaddress{\street{Acad. Vernadsky blvd}, \city{Kyiv}, \postcode{03142}, \country{Ukraine}}}




\abstract{We study the correlated electronic structure and local spin behaviour of the copper-substituted lead–vanadium bromine apatite Pb$_9$Cu(VO$_4$)$_6$Br$_2$ using DFT+DMFT with a two-orbital Cu-centred low-energy model.  Simulations are done for several temperatures (20, 60, 100 K) and a broad range of band fillings 2.46 $\leq$ n $\leq$ 3.54. We find that the present compound stays metallic even once correlations are treated dynamically around the stoichiometric filling (n $\simeq$ 3). Away from n $\simeq$ 3, both hole and electron doping drive the system toward non-Fermi-liquid behaviour, and spectral weight is transferred from the low-energy peak into upper and lower Hubbard-like features. By analysing the low-frequency self-energy exponent and the dynamical part of the local spin susceptibility, we identify a narrow window of enhanced spin fluctuations on the slightly hole-doped side (n $\simeq$ 2.94), i.e. a spin-freezing-crossover regime of the kind reported in the literature for multiorbital Hund metals. This places Pb$_9$Cu(VO$_4$)$_6$Br$_2$ among the promising members of the Cu-substituted apatite family. }




\maketitle

\section{Introduction}\label{sec1}

Copper-substituted lead apatites briefly drew a great deal of attention after early reports of high-temperature superconductivity, which were later refuted on experimental grounds \cite{korean_apatites_2, sukhenko2024}. Those claims nonetheless triggered a series of theoretical studies that pointed out an appealing combination of features in this structural family: flat-band-like electronic states, strong correlations, and competing magnetic tendencies. Despite the fact that these characteristics were not realised simultaneously in the specific original composition, the broader apatite-type Cu compounds remain an interesting platform in which such physics could, in principle, be engineered.

In a recent preprint \cite{ladder}, using thermodynamic stability and symmetry robustness, we have discovered an apatite-type compound Pb$_9$Cu(VO$_4$)$_6$Br$_2$ which should remain metallic even without doping, which is the key feature distinguishing it from its other apatite peers. In this paper, we check if this conclusion holds when the strong correlations are treated properly - within the higher-level theory such as DMFT.


However, rather than simply testing robustness of the metallic solution and analysing  quasiparticle renormalisation, we will also shed light on the behaviour of the local moments at the copper site. 

When describing the behaviour of local spins in multiorbital systems with Hund coupling, one might imagine two extremes: either a coherent Fermi-liquid metal, in which local spins are rapidly fluctuating, or an incoherent “spin-frozen” metal, in which long-lived local moments appear and the self-energy acquires a non-Fermi-liquid form, leading to a breakdown of quasiparticle description \cite{werner-2008}. The most interesting scenario lies right in between: when spins are neither short-time fluid nor fully static. This peculiar \textit{spin-freezing-crossover} (SFC) regime has been linked to normal state of unconventional superconductors: it has been shown that these spin fluctuations might provide a glue for the spin-triplet pairing \cite{werner-2015}. It has been further shown by P.~Werner and colleagues that the SFC regime lies above the superconducting dome of the phase diagram of various important superconductor families: iron pnictides \cite{werner-2012}, strontium ruthenate \cite{werner-SrRu}, nickelates \cite{werner-Ni}, and even cuprates, whose multi-site single-orbital model may be transformed into a single-site multi-orbital one with the emergent Hund coupling \cite{werner-Cu}. Thus, while the spin-freezing crossover does not constitute a definitive proof of superconductivity, it may serve as an early indicator of it. In fact, when it comes to copper-substituted lead apatites, this framework has already been applied to the controversial LK-99 \cite{correlated}, but the latter was found to be insulating if stoichiometric, and deeply spin-frozen under doping. Here, having first discovered an arguably more promising member of the family, we shall also include the local spin in our DMFT study.

This paper aims to answer the following questions:

\begin{enumerate}
    \item Does the compound remain metallic once strong correlations are treated properly?
    \item What are the spectral fingerprints (quasiparticle renormalisation, Hubbard bands)?
    \item Do local moments freeze or fluctuate, and is there a spin‑freezing crossover (Hund‑metal physics) on the electron- or hole-doped side?
\end{enumerate}






\section{Theory and Methods}\label{sec11}

\subsection{DMFT} \label{ssec:dmft}


For a non-interacting one-electron band $\varepsilon_{\mathbf{k}}$, the spectral function

\begin{equation}
    A_{0} (\omega, \mathbf{k}) = 2\pi \delta (\omega - \varepsilon_{\mathbf{k}}),
\end{equation}

where $\mathbf{k}$ is the wave vector (momentum), and $\omega$ is the excitation frequency (i.e. energy depth). Here the self-energy $\Sigma = 0$. 

In case of DFT+U (a static, Hartree-Fock-like correction), the on-site Coulomb repulsion is treated as a static mean field, therefore the self-energy is frequency-independent:

\begin{equation}
    A_{DFT+U} (\omega, \mathbf{k}) = 2\pi \delta \left( \omega - \varepsilon_{\mathbf{k}} - \Sigma_0 \right)
\end{equation}

It does introduce the energy shift, but the quasiparticle lifetime is infinite. Metal-to-insulator transition (MIT) can only be achieved by breaking the symmetry, such as by introducing antiferromagnetic supercells \cite{eva}.

Instead, dynamical mean-field theory \cite{dmft-1} (DMFT) transforms the local lattice problem (such as the Hubbard model) into the impurity problem, such as Anderson's. There, the Green's function of the lattice, the spectral function and the quasiparticle weight are:

\begin{align}
    G(\omega, \mathbf{k}) = \left ( \omega - \mu  - \varepsilon_{\mathbf{k}} - \Sigma(\omega) \right)^{-1} \nonumber \\
    A_{\text{DMFT}} (\omega, \mathbf{k}) = - \frac{1}{\pi} \text{Im} G  (\omega, \mathbf{k})  \nonumber \\
    Z = \left( 1 - \left. \frac{\partial \mathrm{Re}\,\Sigma(\omega)}{\partial \omega} \right|_{\omega=0} \right)^{-1}
\end{align}

The crucial difference to DFT+$U$ is the \emph{frequency dependence} of the (local) self–energy: it allows spectral weight to be redistributed from the low–energy quasiparticle peak into incoherent Hubbard–like features at higher energies. As correlations increase, $Z \to 0$ and $A(\omega{=}0)\to 0$, signalling a Mott–like metal–insulator transition.

At finite temperature it is both customary and convenient to work in imaginary time $\tau \in [0,\beta]$, where $\beta = 1/k_B T$. Antiperiodicity of fermionic correlators dictates that their Fourier series translate to fermionic Matsubara frequencies:
\begin{equation}
    \omega_n = (2n+1)\frac{\pi}{\beta}, \ \ \ \ n \in Z.
\end{equation}
This imaginary-axis formulation helps avoid singularities that may creep in with real frequencies and makes $G$ well-behaved overall. Low $n$ range represents low-energy and long-time physics \cite{Bruus2004}. Let's also note that in a regular metal, the imaginary part of the Matsubara self energy Im$\Sigma _{\omega_{n} \to 0}\to 0$ linearly. This is because for small $\omega_n$:
\begin{equation}
    \text{Im} \Sigma (i \omega_n) \to - \left ( 1 - 1/Z \right ) \omega_n,
\end{equation}

while a constant or diverging $\Im\Sigma(i\omega_n)$ at low $\omega_n$ indicates a breakdown of Fermi–liquid behaviour and, in DMFT, often a Mott–like state, since
\begin{equation}
  G(i\omega_n,\mathbf k) = \frac{1}{i\omega_n - \mu - \varepsilon_{\mathbf k} - \Sigma(i\omega_n)}
\end{equation}
is then suppressed at low frequency.

Lastly, from the fact that the local Green's function can be expressed in imaginary time as:

\begin{equation}
    G (\tau) = \int^{+\infty}_{-\infty} d \omega \frac{e^{-\omega \tau}}{1 - e^{- \beta \omega}} A (\omega),
\end{equation}

it can be shown that $G ( \beta /2) \propto A (\omega=0)$. Thus the value $G ( \beta /2)$, directly accessible from a calculation, may serve as a probe of metallicity of a system.

\subsection{Spin Freezing signatures}

In a normal metal, the scattering is dominated by two-electron processes - hence Im~$\Sigma (\omega) \sim \omega^2$, which in Matsubara terms translates to Im~$\Sigma (i\omega_n) \sim \omega_n$. In a non-Fermi liquid with frozen local moments, the scattering is elastic and Im$\Sigma (i\omega{_n} \to 0) \sim \Gamma$, where $\Gamma$ is some constant scattering rate. To capture the \textit{intermediate} state, Im $\Sigma$ is fitted as \cite{werner-2008} 
\begin{equation} \label{eq:exp}
    \text{Im} \Sigma (i\omega_n \to 0) = C + A(\omega_{n})^{\gamma},
\end{equation}
and the spin-freezing crossover is identified by $\gamma \approx 0.5$.

More directly, local spin fluctuations associated with this transition can be characterised by a spike in the dynamical spin susceptibility,

\begin{equation}\label{eq:chi}
    \Delta \chi = \int_{0}^{1/T} d \tau \left( \langle S_{z}(\tau)S_z(0) \rangle   - \langle S_{z}(1/2T)S_z(0) \rangle \right )
\end{equation}

where $\left< S_{z}S_{z} \right>$ is a spin correlation, and the second term in \ref{eq:chi} represents long-term moments that are subtracted. 


\subsection{Computational Details}

For the density functional part, Quantum ESPRESSO package \cite{QE-2009,QE-2017,QE-2020} was used, in conjunction with Meta-GGA r$^2$SCAN functional \cite{r2scan} and norm-conserving pseudopotentials \cite{oncv}. For structural optimisations, A 4×4×6 k-point mesh was used for structural optimisations, and a 6×6×8 mesh - for wannierisation and density of states calculations. The plane-wave energy cutoff was set at 953 eV (70 Ry). Wannier functions were built with WANNIER90 \cite{wannier90}. Correlation parameters $U$ and $J$ were determined from a cRPA calcultaion performed with RESPACK \cite{respack, wan2respack}. For the Cu site of Pb$_9$Cu(VO$_4$)$_6$Br$_2$ these amounted to $U = 5.07$ eV, $J = 0.46$ eV.

For the DMFT part, the single-particle Hamiltonian built with DFT+Wannier was augmented by the Hubbard-Kanamori Hamiltonian: more in section \ref{ssec:lem}. DMFT calculations were performed within TRIQS/solid\_dmft package \cite{triqs1, triqs3, soliddmft} using continuous-time quantum Monte Carlo (QMC) solver in the hybridization
expansion \cite{triqs2}. Analytic continuation was performed with the maximum-entropy (MaxEnt) algorithm \cite{maxent}.

\subsection{Low-energy model} \label{ssec:lem}




\begin{figure}[H]
\centering
\includegraphics[width=1.0\textwidth]{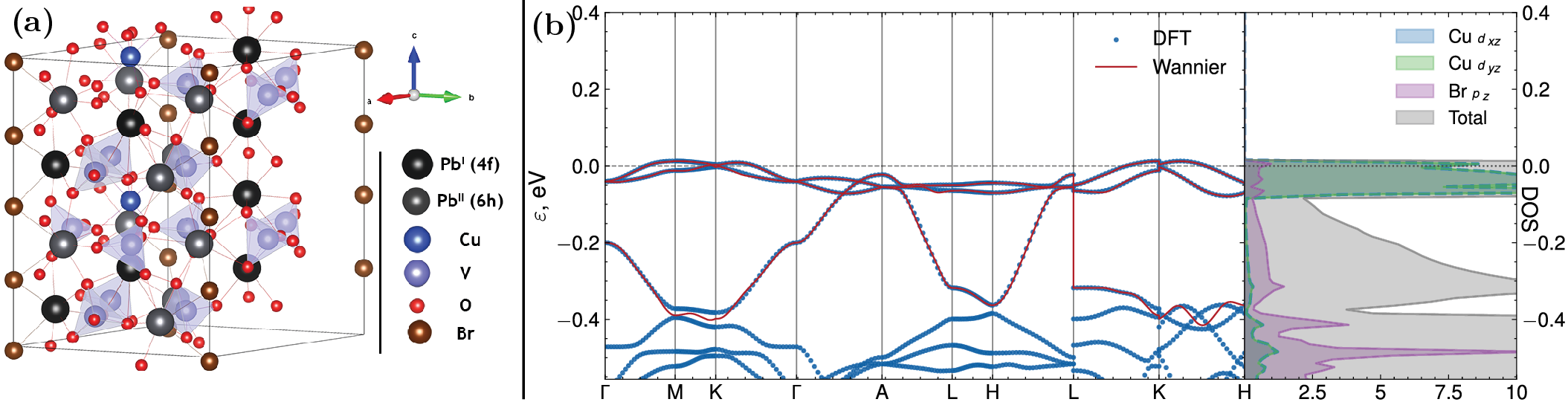}
\caption{a) A schematic view of the crystal structure of Pb$_9$Cu(VO$_4$)$_6$Br$_2$. b) Near-Fermi DFT band structure and densities of states corresponding to Wannier projections.}\label{fig:dftwannier}
\end{figure}

The near-Fermi band structure is presented on Figure \ref{fig:dftwannier}, b.

A three-band Hamiltonian was constructed with Wannier functions to reproduce the low-energy physics. Wannier projections were built with Cu $3d_{xz}$, Cu$3d_{yz}$ and Br $4p_{z}$ orbitals on the merit of them being the largest individual contributors to these three bands within the projected density of states. For the interacting part of the general Hamiltonian, only Cu $3d_{xz}$, $3d_{yz}$ orbitals were considered, thus the DMFT was solved for a two-orbital model. Interacting Hamiltonian was of Hubbard-Kanamori type:

\begin{align}
H_{\text{loc}} =\;&
\sum_{m \in \{a,b\}} \sum_{\sigma} 
    \varepsilon_m\, n_{m\sigma} + U \sum_{m \in \{a,b\}}
    n_{m\uparrow}\, n_{m\downarrow}
\nonumber \\
&+ U' \sum_{\sigma} 
    n_{a\sigma}\, n_{b\bar\sigma}
\;+\; (U'-J) \sum_{\sigma} 
    n_{a\sigma}\, n_{b\sigma}
\nonumber \\
&- J \sum_{\sigma}
\Bigl(
    c^{\dagger}_{a\sigma}\, c_{b\sigma}\,
    c^{\dagger}_{b\bar\sigma}\, c_{a\bar\sigma}
    +
    c^{\dagger}_{b\sigma}\, c_{a\sigma}\,
    c^{\dagger}_{a\bar\sigma}\, c_{b\bar\sigma}
\Bigr)
\nonumber \\
&+ J 
\Bigl(
    c^{\dagger}_{a\uparrow}\, c^{\dagger}_{a\downarrow}\,
    c_{b\downarrow}\, c_{b\uparrow}
    +
    c^{\dagger}_{b\uparrow}\, c^{\dagger}_{b\downarrow}\,
    c_{a\downarrow}\, c_{a\uparrow}
\Bigr)\,,
\end{align}

Dynamical mean-field theory calculations were performed for temperatures of 20, 60 and 100 K. To elucidate the effect of doping, the filling parameter $n$ was also varied. For our two-orbital model, a stoichiometric case corresponds to three electrons in two orbitals ($n = 3$). We studied a range of $n \in [2.46, 3.54]$, thus representing doping from 18 \% holes to 18 \% electrons. Should the experimental synthesis of these samples be performed, it appears that the simplest way to dope them would be the substitution of VO$_4$ for TiO$_4$ or CrO$_4$ \cite{cr04, tio4}.

\section{Results}

\subsection{Metallicity and fermi-liquidness}

\begin{figure}[H]
\centering
\includegraphics[width=0.9\textwidth]{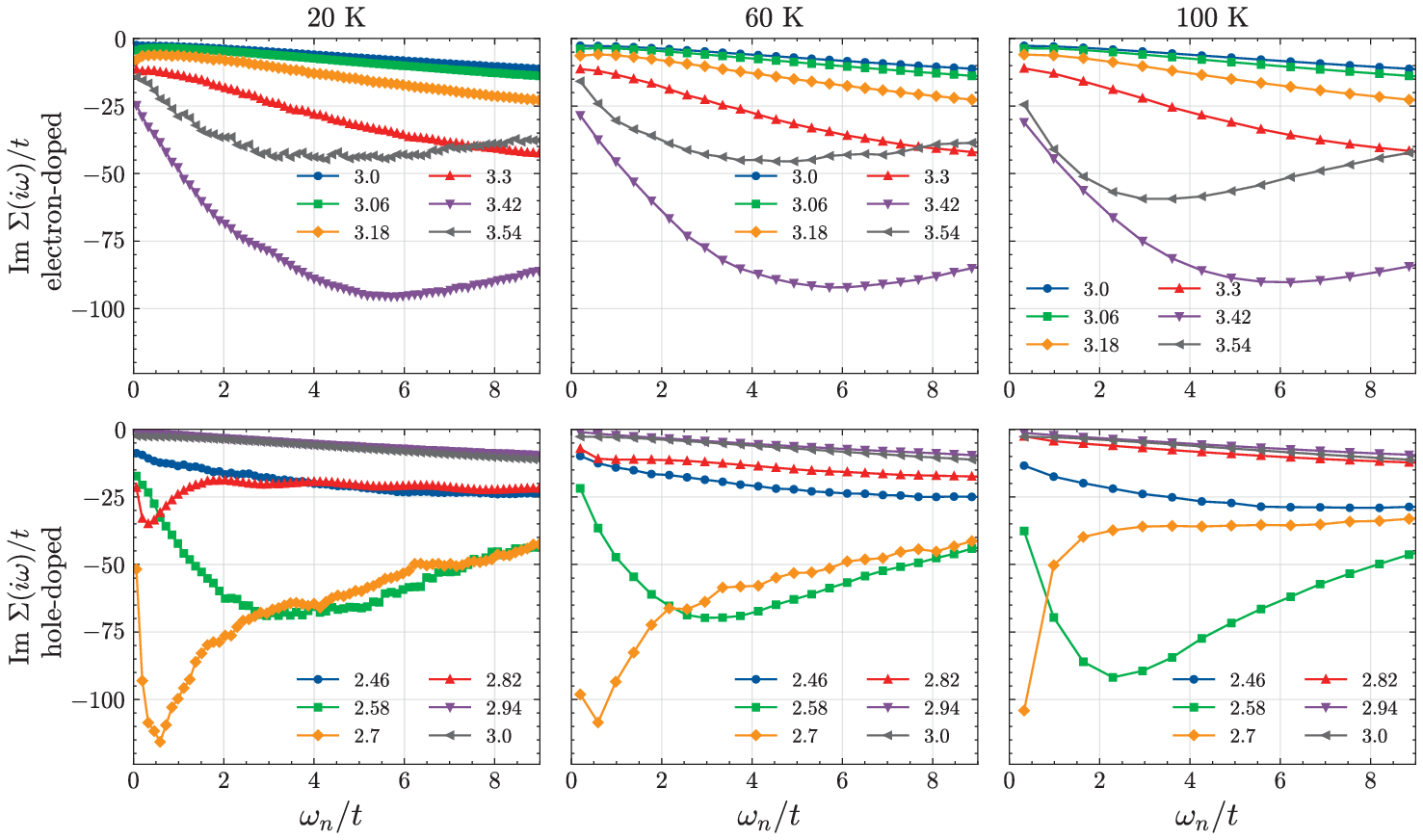}
\caption{Imaginary part of the self-energy as a function of Matsubara frequencies for various doping levels and temperatures.  }\label{fig:imsigma}
\end{figure}

Low-frequency behaviour of the imaginary part of the Matsubara self-energy is presented at Figure \ref{fig:imsigma}. As described in \ref{ssec:dmft}, judging by this value, we can estimate how close to Fermi liquid our system is. 

The values of Im $\Sigma$ were averaged over two spins and two orbitals. The most FL-like behaviour is observed around $n=3$, particularly from $n = 2.94$ to 3.06. For other doping regimes, Im $\Sigma$ approaches a constant for $\omega_n \to 0$, indicating severe deviations from the Fermi liquid, indicating a breakdown of quasiparticle description. The closest to divergence curve is found at $n = 2.7$, indicating an approach to the metal-to-insulator transition.

\begin{figure}[H]
\centering
\includegraphics[width=0.9\textwidth]{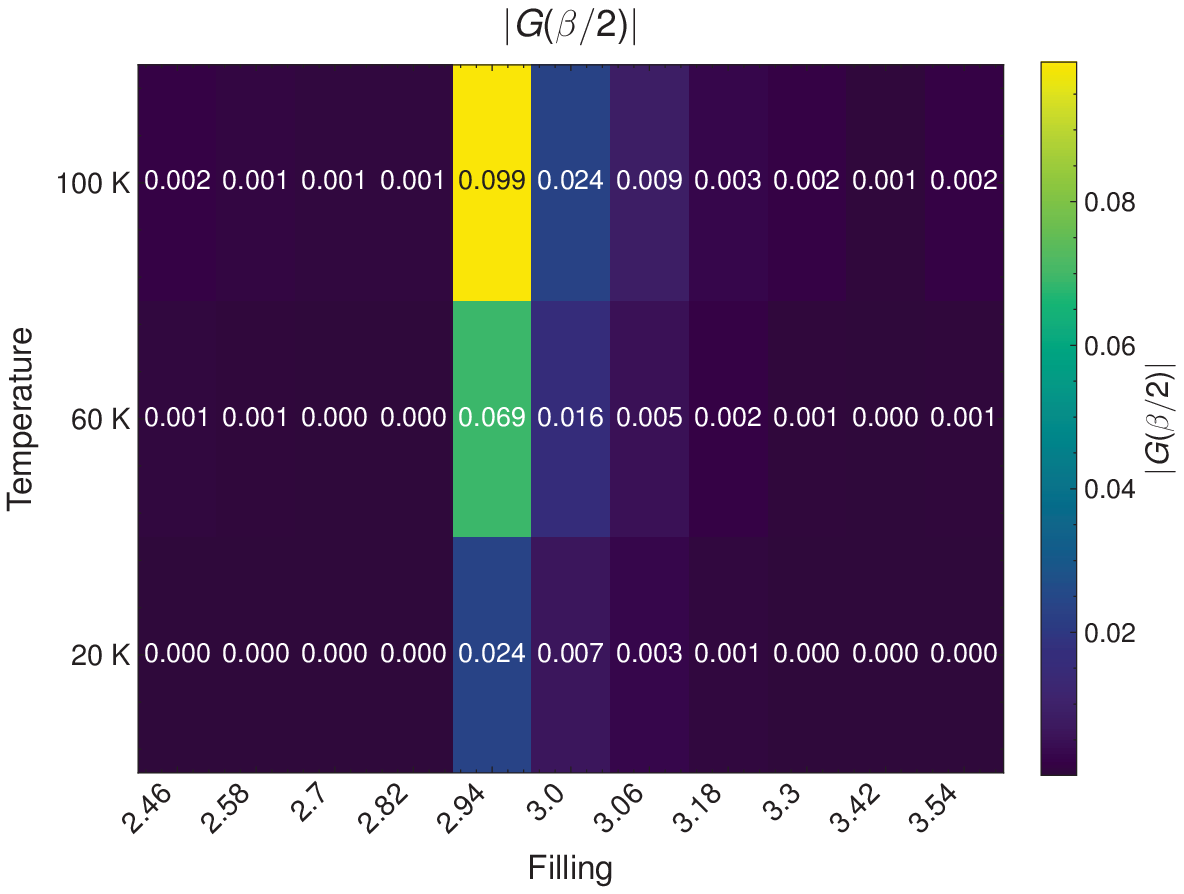}
\caption{Absolute value of the local Green's function at half the inverse temperature - a probe of the spectral weight at $E_F$ - for different fillings and temperatures.  }\label{fig:colormap}
\end{figure}

To estimate a range of robustly metallic solutions, we plot a $\left | G (\beta/2) \right|$  (Figure \ref{fig:colormap}).

As discussed in \ref{ssec:dmft}, this value is proportional to the spectral function at $E_F$. The heaviest spectral weight is observed for $n= 2.94$, followed by stoichiometric $n=3$. For hole-doped side, it drops brusquely below $n = 2.94$, while for electron-doped side the decrease is more gradual.

\begin{figure}[H]
\centering
\includegraphics[width=0.9\textwidth]{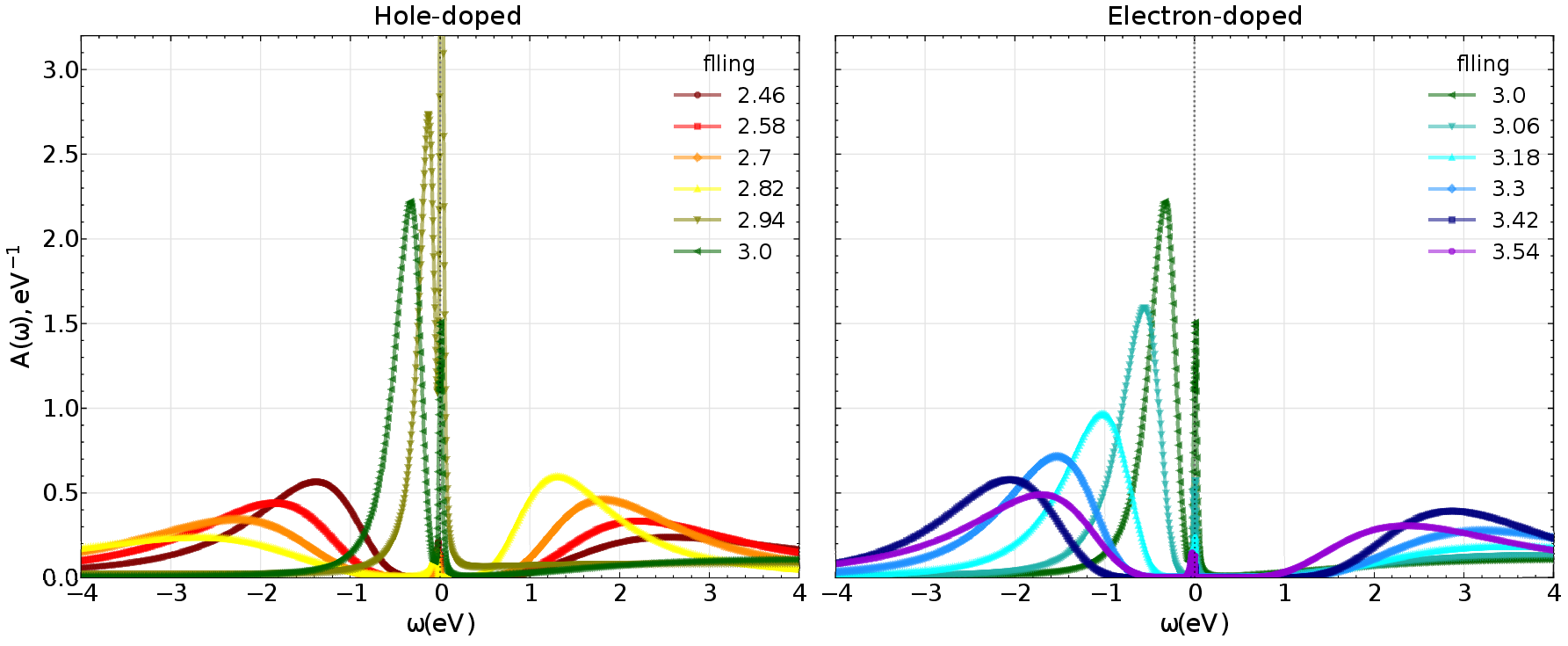}
\caption{Analytically continued spectral function for T = 20 K and various fillings.}\label{fig:Aw}
\end{figure}

To elucidate the spectral function explicitly, we performed the analytic continuation of impurity Green's function to the real frequency axis using Maximum Entropy algorithm \cite{maxent}. The result for T = 20 K is presented at the Figure \ref{fig:Aw}

In agreement with $\left | G (\beta/2) \right|$ estimate, $A (\omega = 0) $ demonstrates the highest peak for $n = 2.94$, followed by $n= 3.0$. 

Curiously, $A (\omega = 0)$ never quite reaches zero, that is, never drops below the MaxEnt error of 0.01 units, with the lowest value observed for $n=2.7$, complementing the divergence at Figure \ref{fig:imsigma}.

For fillings close to n $\approx 3$ the spectral function shows a narrow quasiparticle peak located at $E_F$, also in agreement with $ | G (\beta/2)|$ and Im $\Sigma$ estimates. In the undoped case, the broad upper Hubbard band peaks around $U - J = 4.61$ eV. Doping away from $n = 3$, the lower Hubbard band slides away from $E_F$ and broadens, while the QP peak's weight gets consumed mostly by the upper Hubbard band that grows in intensity.

These results are quite remarkable if put into context: in its time, a major reason for the failure of LK-99 was its insulating character, revealed both by experimental and theoretical studies. Physics behind this is well explained in \cite{georgescu}: due to the large distance between isolated copper sites, it becomes easy for the system to lower the energy by opening a gap, which occurs through a Jan-Teller structural distortion. At the same time, DMFT studies \cite{mott-or-charge, correlated} confirmed its insulating state (without doping). Notably, with a more sophisticated eDMFT multi-band approach, Kim et al. \cite{Kim2024} expanded those conclusions and did find a metallic state in an undoped LK-99 at \textit{some} U\&J values, albeit not at its natural U\&J.  
All the while, using symmetry-broken DFT, in \cite{ladder} we showed that Pb$_9$Cu(VO$_4$)$_6$Br$_2$, which shares the same structural type, is likely to survive the structural distortion and remain metallic. Here, DMFT results confirm this conclusion. The physical reason for this is copper charge is not as isolated anymore: as seen on Fig. \ref{fig:dftwannier}, b, two flat Cu $3d_{xz /yz}$ bands are in touch and cross with a dispersive Br $4p_z$ band, whereas in real space, Br$^-$ provide a “charge bridge” thanks to their larger ionic radii. Note that the correlation regime parameter $U/t$ which would be $\approx 33$, becomes $\approx 11.5$, which is still very strongly correlated, but noticeably milder.

\subsection{Spin Freezing Crossover?}

\begin{figure}[H]
\centering
\includegraphics[width=1.0\textwidth]{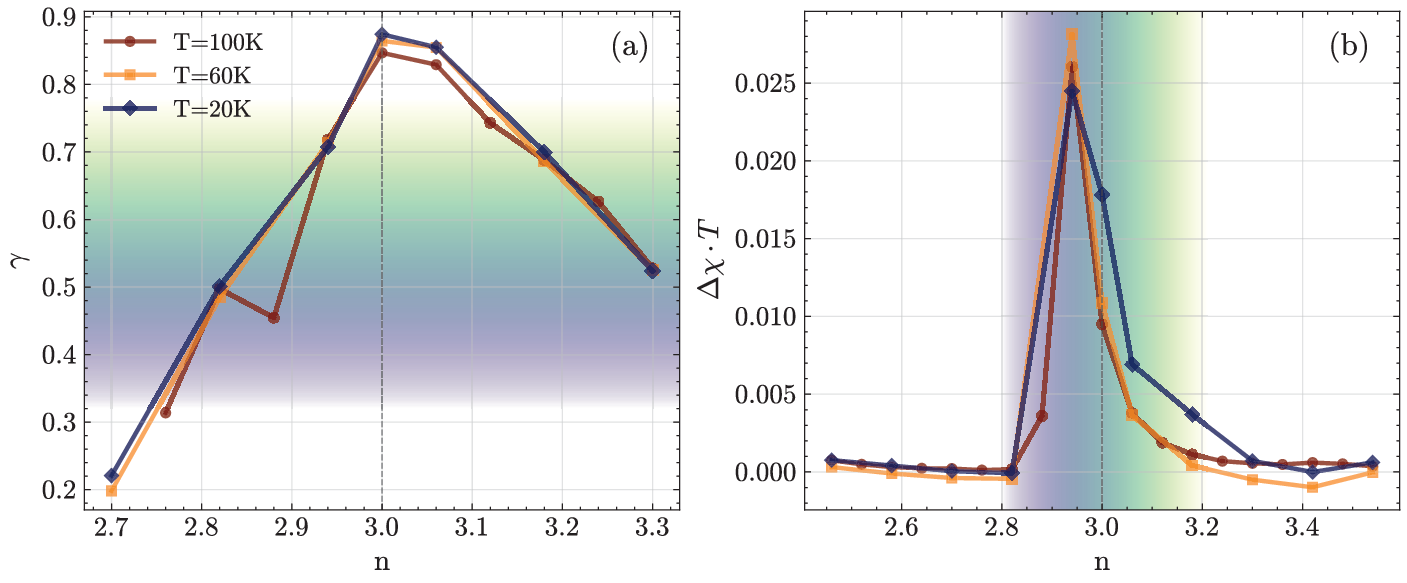}
\caption{Non-Fermi-liquid exponent factor $\gamma$ and the local spin susceptibility $\chi$. Highlighted in gradient are the regions of suspected spin-freezing crossover as suggested by the plotted values.}\label{fig:gammachi}
\end{figure}

Having drawn the spectral portrait, let us examine the local spin behaviour and situate our compound within “spin-frozen versus regular metal” space.

Following the procedure in \cite{werner-2008}, we fit the Im $\Sigma (\omega_n)$ (Fig. \ref{fig:imsigma} by the exponential function \ref{eq:exp}. The exponential factor $\gamma$ is plotted at the Figure \ref{fig:gammachi}, a. We observe $\gamma$ is the closest to FL-like value of 1 at $n = 3$, while with doping it drops towards zero, indicating the breakdown of the quasiparticle description, freezing of moments and approach to the insulating transition. 

Next, at Fig. \ref{fig:gammachi}, b, we plot the dynamical contribution to the spin susceptibility \ref{eq:chi}. It displays a sharp peak at $n = 2.94$, which points strongly to an increase in spin fluctuations. The value of $n = 2.94$ is also the value at which, judging by $ | G (\beta/2)|$ and $A(\omega=0)$, the spectral weight at $E_F$ is the largest.   

According to the spin freezing theory, the “sweet spot” of the crossover corresponds to $\gamma$ between 1 and 0 around $\gamma \approx 0.5$. For the filling $n = 2.94$ that corresponds to maximum $\chi_{dyn}$, the exponential factor $\gamma = 0.7$, indicating an approach to the crossover region.

\begin{figure}[H]
\centering
\includegraphics[width=0.6\textwidth]{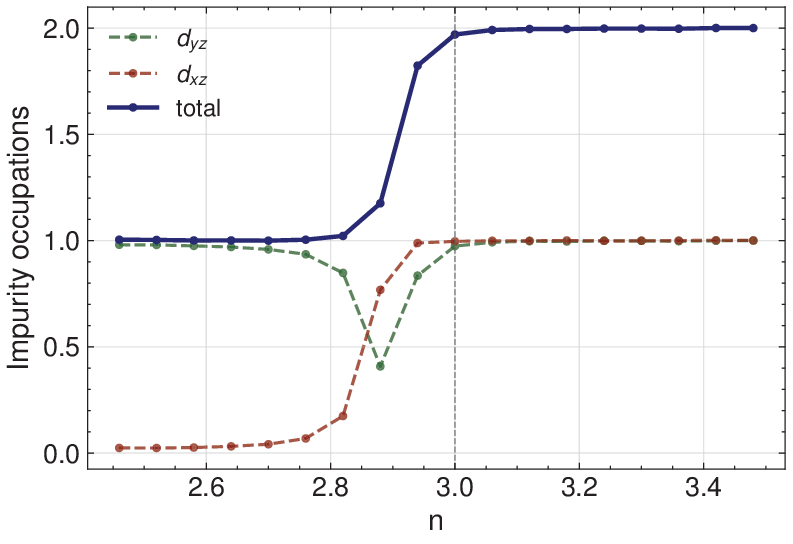}
\caption{Occupation numbers of the correlated orbital subset.}\label{fig:occs}
\end{figure}

Physically, the spike in spin fluctuations corresponds to the transition in the occupation of the orbital subset from 1 at strong hole doping to 2 at strong electron doping. This crossover is accompanied by the Hund coupling that is trying to establish the high-spin configuration. This evolution is demonstrated at Figure \ref{fig:occs}. Spin-frozen regime is found at both terminal ends of the doping range. Note that the DMFT impurity occupancy refers to the two Cu-centred Wannier orbitals ($d_{xz}$, $d_{yz}$) in the correlated subspace. This subspace does not include the Br $4p$, O $2p$ that also carry some Cu $d$ weight. Still, trends of the DMFT occupancy with filling and temperature are meaningful, because the chemical potential is shifted self-consistently.






\section{Conclusion}\label{sec13}

In this paper, we explored the properties of a promising strongly correlated material, Pb$_9$Cu(VO$_4$)$_6$Br$_2$, by means of the dynamical mean-field theory. We have demonstrated that it does not experience a Mott transition at zero doping, staying metallic. 

Additionally, the behaviour of spin at the Cu site was examined. At doping of $n = 2.94$, corresponding to 2\% hole doping, a spike in spin fluctuations was recorded, evidenced by a sharp peak of the dynamical local spin susceptibility. It stems from the transition from a single-electron to a two-electron occupation of the correlated band subset, and indicates the presence of the spin freezing crossover.

Within the spin freezing theory of superconductivity, local spin fluctuations at the crossover edge are linked to the pairing mechanism for spin-triplet superconductivity. Thus, discovery of this regime in our work may serve as an inspiration for synthesis and characterisation of the lead-copper-vanadium-bromine apatite. 

Finally, it has to be stated that the magnetic order was not subject of this study, since the single-site DMFT cannot resolve the long-range order. This, as well as the extension of the two-band model to other bands might merit consideration in a future study.

\backmatter





\bmhead{Acknowledgements}

This work was supported by the National Research Foundation of
Ukraine (Grant No. 2023.03/0242).









\bibliography{sn-bibliography}

\end{document}